\documentclass[aip,jcp,preprint,amsmath,amssymb,nobalancelastpage]{revtex4-2}
\newcommand{\vv}[1]{\mathbf{#1}}
\renewcommand{\d}[1]{\ensuremath{\operatorname{d}\!{#1}}}
\usepackage{graphicx}
\usepackage[suffix=]{epstopdf}
\usepackage{natmove}

\usepackage{xr}
\makeatletter
\newcommand*{\addFileDependency}[1]{% argument=file name and extension
  \typeout{(#1)}
  \@addtofilelist{#1}
  \IfFileExists{#1}{}{\typeout{No file #1.}}
}
\makeatother
\newcommand*{\myexternaldocument}[1]{%
    \externaldocument{#1}%
    \addFileDependency{#1.tex}%
    \addFileDependency{#1.aux}%
}
\myexternaldocument{si}

\begin{document}

\title{Designing binary mixtures of colloidal particles with simple interactions that assemble complex crystals}
\author{Beth Hsiao-Yen Wei}
\thanks{These authors contributed equally.}
\affiliation{Artie McFerrin Department of Chemical Engineering, Texas A\&M University, College Station}

\author{C. Levi Petix}
\thanks{These authors contributed equally.}
\affiliation{Department of Chemical Engineering, Auburn University, Auburn, AL 36849, USA}

\author{Qizan Chen}
\affiliation{Artie McFerrin Department of Chemical Engineering, Texas A\&M University, College Station}

\author{Michael P. Howard}
\email{mphoward@auburn.edu}
\affiliation{Department of Chemical Engineering, Auburn University, Auburn, AL 36849, USA}

\author{Jeetain Mittal}
\email{jeetain@tamu.edu}
\affiliation{Artie McFerrin Department of Chemical Engineering, Texas A\&M University, College Station}
\affiliation{Department of Chemistry, Texas A\&M University, College Station}
\affiliation{Interdisciplinary Graduate Program in Genetics and Genomics, Texas A\&M University, College Station}

\begin{abstract}
Computational methods for designing interactions between colloidal particles that induce self-assembly have received much attention for their promise to discover tailored materials. However, it often remains a challenge to translate computationally designed interactions to experiments because they may have features that are too complex, or even infeasible, to physically realize. Toward bridging this gap, we leverage relative-entropy minimization to design pair potentials for binary mixtures of colloidal particles that assemble crystal superlattices. We reduce the dimensionality and extent of the interaction design space by enforcing constraints on the form and parametrization of the pair potentials that are physically motivated by DNA-functionalized nanoparticles. We show that several two- and three-dimensional lattices, including honeycomb and cubic diamond, can be assembled using simple interactions despite their complex structures. We also find that the initial conditions used for the designed parameters as well as the assembly protocol play important roles in determining the outcome and success of the assembly process.
\end{abstract}

\maketitle

\section{Introduction}
The self-assembly of colloidal particles into crystal superlattices has important applications in the fabrication of optical \cite{doi:10.1021/acs.chemrev.6b00196}, catalytic \cite{doi:10.1021/ja3097527} and plasmonic materials \cite{doi:10.1073/pnas.1422649112}. Particle properties such as shape \cite{Lee2022}, patchiness \cite{Glotzer2007,C9CS00804G}, size \cite{doi:10.1021/nl300679e}, charge \cite{PhysRevLett.96.018303,Leunissen2005} and surface functionalization \cite{doi:10.1021/ja108948z} can be engineered to control the structure of the superlattice and its corresponding properties \cite{Wang2020, doi:10.1126/science.adl2733}. Characterizing the different superlattices that are produced from such a vast design space using traditional ``forward'' experimentation can be expensive and slow, motivating the use of ``inverse'' methods that efficiently identify particles that yield desired structures. Inverse-design methods for self-assembly typically consist of a loss function that measures how different the produced structure is from the desired one and an optimization strategy that minimizes the loss function with respect to the particle properties that can be varied \cite{sherman:jcp:2020}. These methods include both approaches physically grounded in statistical mechanics \cite{PhysRevLett.95.228301, C0SM01205J, PhysRevE.75.031403, 10.1063/1.4790634}, such as free-energy landscape engineering \cite{C7ME00077D, C9SM01500K}, relative-entropy minimization \cite{doi:10.1021/acs.jpcb.7b11841, 10.1063/1.4981796, 10.1063/1.4962754, 10.1063/1.5021648, C5SM01832C, C6SM02718K, 10.1063/1.5088604}, and digital alchemy \cite{doi:10.1021/acsnano.5b04181, D2SM01593E}, as well as data-driven strategies that exploit machine learning\cite{doi:10.1126/sciadv.aax9324,doi:10.1126/sciadv.abj6731,10.1063/1.5111492,2410.22111}.

Inverse design has had considerable computational success \cite{sherman:jcp:2020, Dijkstra2021}, but translation to experiment is far more limited \cite{doi:10.1073/pnas.1316533110, 2310.10995}. One major challenge is that pairwise potential energy functions, representing the effective interaction between two particles, are typically designed computationally, but creating particles that will produce these interactions can be difficult if the pair potentials do not have physically interpretable parameters. Pair potentials having no experimental realization may even be designed if physical requirements are not enforced in the optimization. For example, relative-entropy minimization has been successfully applied many times to design pair potentials represented using Akima splines \cite{10.1063/1.4962754, 10.1063/1.5021648, 10.1063/1.4981796, doi:10.1021/acs.jpcb.7b11841, 10.1063/1.5088604, 10.1063/1.5063802}. Splines have great flexibility in the functions they can represent and so are useful for finding a feasible design, but this flexibility may produce features such as high-frequency oscillations or plateaus that do not readily map to a known particle surface chemistry. Spline potentials can be constrained \cite{doi:10.1021/acs.jpcb.7b11841} or filtered \cite{10.1063/1.5063802} to remove some of these features, but the resulting interactions  are still typically challenging to realize. Using multiple types of particles has also been proposed as a strategy to simplify interaction complexity at the expense of compositional complexity \cite{Mahynski2019}; however, spline potentials designed for binary mixtures of particles using relative-entropy minimization were not necessarily simpler than those designed for equivalent one-component assemblies \cite{10.1063/1.5021648}.

We postulate that this challenge is not caused by a nonexistence of experimentally producible designs. The loss function guiding the design may have local minima and/or regions with low curvature near a global minimum, suggesting that other, simpler interactions may produce acceptable or near-optimal structures. Indeed, forward modeling has shown that colloidal particles interacting through simple, often experimentally-motivated, pair potentials can assemble a variety of structures \cite{doi:10.1021/acs.jpclett.7b02237, PhysRevE.98.052601,doi:10.1073/pnas.1504677112}, but these interactions may not be selected by inverse methods if they are less optimal than other, more complex interactions that are admissible. Inverse-design methods can be forced to consider only these simpler interactions by designing physically motivated pair potentials with parameters that are directly related to experimentally controllable variables \cite{bedolla2024, 10.1063/5.0048812, C9SM02426C}. The tradeoff is that it may now be more difficult, or even impossible, to produce the desired structure than with a spline potential because the particle interactions are more constrained.

At the same time, DNA-functionalized particles (DFPs) are known to be a versatile material platform of particular interest for colloidal self-assembly. Simple pair potentials \cite{doi:10.1021/la701166p, Auyeung2014, PhysRevLett.89.148303}, including Fermi--Jagla\cite{doi:10.1021/acs.langmuir.7b02835, C8SM00989A} and Lennard-Jones-$nm$\cite{Mao} potentials, have been developed to capture the effective interactions between DFPs, such as hybridization-driven attraction and repulsion due to chain overlaps, and have been experimentally validated \cite{10.1002/smll.200700357, doi:10.1126/science.1259762, Mirkin1996, C7SM01722G, doi:10.1126/science.1210493, mi13071102}.
While experimental self-assembly of single-component crystals has mostly been limited to simple close-packed structures \cite{doi:10.1073/pnas.0900630106,doi:10.1126/science.1210493,doi:10.1021/acs.nanolett.5b02129}, binary mixtures of DFPs offer a richer design space because stoichiometry and differences in particle interactions can be exploited to stabilize structures that would be inaccessible with only a single type of DFP \cite{doi:10.1126/science.1210493, Wang2015}.

In this work, we use relative-entropy minimization to optimize pair potentials for binary mixtures of colloidal particles under physical constraints that reflect known features of DFP interactions, namely tunable repulsion ranges and attraction strengths, as well as symmetries of the superlattice. Our goal is to show that simple interactions can be designed to assemble complex crystal superlattices---square and honeycomb in two dimensions, body-centered cubic (BCC) and cubic diamond in three dimensions---for binary particle mixtures, even though they should not with only one particle type. We demonstrate that all these superlattices can be successfully designed for using a bulk isochoric temperature cycling  protocol; however, we find that only some designs subsequently reliably self-assemble under bulk isothermal compression. We show that one reason for this behavior may be sensitivity to the initial choice of parameters for optimization, consistent with our postulation about the viability of multiple designs. We additionally probe how our designed interactions perform under dilute conditions typically used in experiments, highlighting the important role of protocol in both the design and assembly processes.

\section{Model and methods} 
We employed the perturbed Lennard-Jones pair potential \cite{weeks:jcp:1971, ashbaughhatch} as a simple model that captures the essential physics of DFP interactions, namely tunable repulsion and attraction; alternative models for DFPs could be used within the same framework if desired. Specifically, the pair potential $u_{ij}$ between two particles of types $i$ and $j$ was
\begin{equation}
u_{ij}(r) =
  \begin{cases}
    u_{ij}^{\rm LJ}(r) + (1-\lambda_{ij}) \varepsilon_{ij},& r \le 2^{1/6} \sigma_{ij} \\
    \lambda_{ij} u_{ij}^{\rm LJ}(r),& {\rm otherwise}
  \end{cases},
  \label{eq:u}
\end{equation}
where $u_{ij}^{\rm LJ}$ is the standard Lennard-Jones potential,
\begin{equation}
  u_{ij}^{\rm LJ}(r) = 4 \varepsilon_{ij} \left[\left(\frac{\sigma_{ij}}{r} \right)^{12} - \left(\frac{\sigma_{ij}}{r}\right)^6\right],
\end{equation}
$r$ is the distance between the centers of the particles, $\varepsilon_{ij}$ is the interaction energy, $\sigma_{ij}$ is the interaction length, and $\lambda_{ij}$ modulates the strength of the attractive tail of $u_{ij}$ independently of its repulsive core. When $\lambda_{ij} = 1$, $u_{ij}$ is the Lennard-Jones potential; when $\lambda_{ij} = 0$, $u_{ij}$ is the purely repulsive Weeks--Chandler--Andersen potential; and when $0 < \lambda_{ij} < 1$, $u_{ij}$ has the same repulsion but less attraction than the Lennard-Jones potential.

Our target crystal superlattices (Table~\ref{tabs1}) are formed from binary mixtures containing particles of types A and B, so in principle, 9 parameters are needed to specify the potential energy function corresponding to eq.~\eqref{eq:u}. However, we were able to reduce the dimensionality and size of this design space using physical insights, which is an advantage of performing inverse design using a physically motivated pair potential. Specifically, we assumed that all particles had the same repulsion strength, coming primarily from the core nanoparticle and grafted DNA chains, at the design temperature, so $\varepsilon_{\rm AA} = \varepsilon_{\rm AB} = \varepsilon_{\rm BB} =\varepsilon$, where $\varepsilon$ can be considered the unit of energy. Further, the symmetry of the 4 lattices we designed (Fig.~\ref{fig1}, Table~\ref{tabs1}) implies that $\sigma_{\rm AA} = \sigma_{\rm BB}$ and $\lambda_{\rm AA} = \lambda_{\rm BB}$. Hence, only $\sigma_{\rm AA}$, $\sigma_{\rm AB}$, $\lambda_{\rm AA}$, and $\lambda_{\rm AB}$ needed to be treated as independent parameters. In preliminary tests, we found $\sigma_{ij}$ to be numerically challenging to optimize concurrently with $\lambda_{ij}$ because eq.~\eqref{eq:u} is a highly nonlinear function of $\sigma_{ij}$. We accordingly considered 4 parameters $\boldsymbol{\theta} = (\sigma^6_{\rm AA}, \sigma^6_{\rm AB}, \lambda_{\rm AA}, \lambda_{\rm AB})$ for optimization. Based on forward modeling and experiments \cite{Mao, 10.1063/1.4900891,D5SM00001G}, we constrained this parameter space to be consistent with $0.5\,\sigma \le \sigma_{ij} \le 2\,\sigma$ and $0 \le \lambda_{ij} \le 1$, where $\sigma$ is the unit of length.

To design the parameters $\boldsymbol{\theta}$ that self-assembled a specific lattice, we minimized the relative entropy \cite{shell:jcp:2008,shell:2016,10.1063/1.4962754}, 
\begin{equation}
S_{\mathrm{rel}}(\boldsymbol{\theta}) = \int \d{\vv{R}}\, p_{0}(\vv{R})\ln \left[\frac{p_{0}(\vv{R})}{p(\vv{R};\boldsymbol{\theta})}\right].
\label{eq:S}
\end{equation}
Here, $p_0(\vv{R})$ is the probability density function to observe a given configuration of particles $\vv{R}$ in a target ensemble of structures, while $p(\vv{R};\boldsymbol\theta)$ is the probability density function to observe the same configuration in a model ensemble where particles interact according to eq.~\eqref{eq:u} with parameters given by $\boldsymbol{\theta}$. The relative entropy is zero when $p$ is identical to $p_0$ and positive otherwise, so $S_{\rm rel}$ is a loss function that can be minimized with respect to $\boldsymbol{\theta}$ to make the model ensemble resemble the target ensemble. For a multicomponent mixture at thermodynamic equilibrium in the canonical ensemble (constant temperature $T$, volume $V$, and number of particles $N_i$ for each type $i$), the gradient of $S_{\rm rel}$ with respect to $\boldsymbol{\theta}$ is \cite{sreenivasan:jcp:2024, petix:jctc:2024}
\begin{align}
\frac{\partial S_{\mathrm{rel}}}{\partial \boldsymbol\theta} = &\sum_{i} \sum_{j} \frac{2 \pi \beta N_i N_j}{V} \nonumber \\
& \times \int_0^{\infty} \d{r}\,r^2\left[g_0^{(i j)}(r)-g^{(i j)}(r ; \boldsymbol\theta)\right] \frac{\partial u_{ij}}{\partial \boldsymbol\theta},
\label{eq:gradS}
\end{align}
where the sums are taken over all particle types, $\beta = 1/(k_{\rm B} T)$ with the Boltzmann constant $k_{\rm B}$, and $g_0^{(ij)}(r)$ and $g^{(ij)}(r;\boldsymbol{\theta})$ are the radial distribution functions for particles of types $i$ and $j$ in the target and model ensembles, respectively. Although eq.~\eqref{eq:S} cannot be readily evaluated for most systems, eq.~\eqref{eq:gradS} can and so can be used to minimize $S_{\rm rel}$ with gradient-based methods. In practice, $g_0^{(ij)}$ is prescribed by the desired target structure, while $g^{(ij)}$ must be simulated for a given $\boldsymbol{\theta}$.

To evaluate $g_0^{(ij)}$, initial configurations that contained at least 1000 total particles were generated at a number density and stoichiometry consistent with the target structure (Table \ref{tabs1}). Each particle was tethered to a lattice site by a harmonic potential $\phi(r) = k r^2/2$,
where $r$ is the distance of the center of the particle from the lattice site and $k$ is a spring constant. A target ensemble of particle configurations was then simulated by drawing an independent random displacement for each particle from a Gaussian distribution with zero mean and variance $(\beta k)^{-1}$. We chose a spring constant $k=1000\,\varepsilon/\sigma^2$, which we found to produce well-defined but numerically integrable peaks in $g_0^{(ij)}$, and used 1000 configurations to compute $g_0^{(ij)}$ up to a distance of $5\,\sigma$ with a bin spacing of $0.05\,\sigma$.

Relative-entropy minimization was carried out with relentless\cite{sreenivasan:jcp:2024} (version 0.1.1 with operations modified to support volume resizing) using HOOMD-blue\cite{anderson:compmatsci:2020} (version 2.9.7). Langevin dynamics simulations were performed to evaluate $g^{(ij)}$ in eq.~\eqref{eq:gradS}. The simulation timestep was $0.001\,\tau$ and the particle friction factor was $0.1\,m/\tau$, where $m$ is the mass of a particle (taken to be the same for both types) and $\tau = \sqrt{m\sigma^2/\varepsilon}$ is the unit of time. The particles were initialized in the target structure, then equilibrated for $3\times10^4\,\tau$ at temperature $T = 3\,\varepsilon/k_{\rm B}$ to attempt to melt the crystal. We found this initialization protocol to be necessary because it was often not possible to place the particles in randomized or fluid-like configurations without overlap for suboptimal values of $\sigma_{ij}$. The temperature was then slowly quenched from $3\,\varepsilon/k_{\rm B}$ to $1\,\varepsilon/k_{\rm B}$ over $2\times 10^4\,\tau$ at a constant rate. The radial distribution function $g^{(ij)}$ was finally sampled every $10\,\tau$ during a $10^4\,\tau$ production period with the same maximum distance and bin spacing as was used for $g_0^{(ij)}$. The parameters $\boldsymbol{\theta}$ were then adjusted to attempt to minimize $S_{\rm rel}$ using gradient descent with a step size chosen for each parameter such that the change in $\sigma_{ij}^6$ was approximately $0.25\,\sigma^6$ and the change in $\lambda_{ij}$ was approximately 0.05 for the first iteration. We initialized $\sigma_{ij}^6$ to be consistent with the position of the first peak in $g_0^{(ij)}$ based on physical knowledge that it relates to the size of the particle and DNA corona, while $\lambda_{\rm AA}$ and $\lambda_{AB}$ were initialized to 0 and 1, respectively, based on previous work \cite{Mahynski2019, Mao}. The optimization was considered converged when the absolute values of all components of the gradient were less than $0.01$ in their respective units or further progress could not be made due to the box constraints on the parameters.

\section{Results and discussion}
\subsection{Initial design}
\label{sec:design}
\begin{figure*}
    \includegraphics{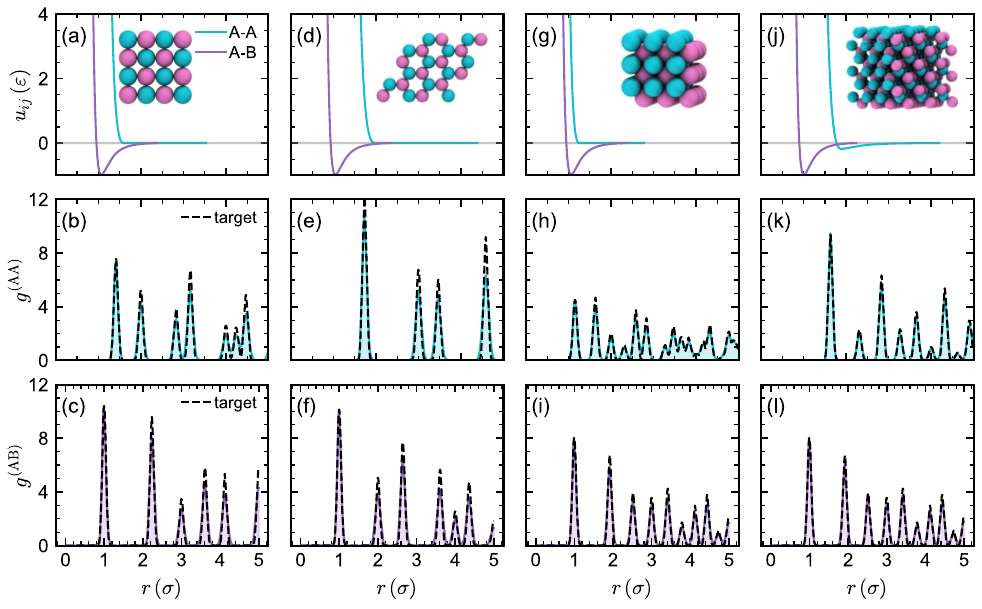}
    \caption{Designed pair potentials and representative snapshots for each crystal (top row) along with $g^{(ij)}$ measured between like (middle row, filled line) and unlike (bottom row, filled line) particles, compared to the target $g_0^{(ij)}$ (dashed line). The four crystal structures are: (a)--(c) square, (d)--(f) honeycomb, (g)--(i) BCC, and (j)--(l) cubic diamond. Snapshots of the target structures (inset) were rendered using VMD (version 1.9.4) \cite{humphrey:jmolgraph:1996}.}
    \label{fig1} 
\end{figure*}

We carried out our inverse-design procedure starting with the square lattice (Fig.~\ref{fig1}a--c). The procedure quickly converged, finding an interaction between like particles that was purely repulsive and an interaction between unlike particles that was attractive; $\lambda_{\rm AA}$ and $\lambda_{\rm AB}$ remained at their initial values of 0 and 1, respectively, throughout the optimization (Fig.~\ref{figs1}). We found that $\sigma_{\rm AA}$ also remained close to its initial value at the first peak in $g_0^{({\rm AA})}$, but $\sigma_{\rm AB}$ somewhat decreased from its initial value at the first peak in $g_0^{({\rm AB})}$. The range of the like-particle repulsions was longer than that of the unlike-particle attractions, which is characteristic of binary DFPs \cite{doi:10.1073/pnas.1109853108,10.1063/1.4900891}. Purely repulsive Akima-spline potentials have been designed to self-assemble the square lattice using both one and two types of particles \cite{10.1063/1.4962754, 10.1063/1.5021648}. Those potentials both have complex features, such as shoulders and plateaus, that our constrained design does not.

Encouraged by this success, we proceeded to design the remaining lattices that we anticipated to be more challenging than square: honeycomb (Fig.~\ref{fig1}d--f) is a low-coordinated lattice, BCC (Fig.~\ref{fig1}g--i) is a three-dimensional lattice, and cubic diamond (Fig.~\ref{fig1}j--l) is a low-coordinated three-dimensional lattice. As for the square lattice, $\sigma_{\rm AA}$ typically remained close to its initial value, set by the first peak in $g_0^{({\rm AA})}$, but $\sigma_{\rm AB}$ tended to decrease for all three structures (Fig.~\ref{figs1}). For honeycomb, $\lambda_{\rm AB}$ departed from its initial value but quickly returned; the same happened with $\lambda_{\rm AA}$ for BCC. Hence, the interactions for both honeycomb and BCC were purely repulsive between like particles and a Lennard-Jones attraction between unlike particles. Interestingly, the cubic diamond was found to have a nonzero $\lambda_{\rm AA}$ when the optimization converged, corresponding to a weak attraction between like particles in addition to a Lennard-Jones attraction between unlike particles. Such attraction can be achieved by mixing two different DNA strands on each particle \cite{Casey2012, doi:10.1021/acs.langmuir.9b03391} or via linker-mediated binding\cite{doi:10.1021/acs.langmuir.9b03391}.

Overall, our designed interactions are consistent with a pattern observed in a forward modeling study of DFPs \cite{Mao}: the denser crystals (square and BCC) were assembled by competing repulsive and attractive interactions that were closer in range than those of the less dense crystals (honeycomb and cubic diamond). For the less dense crystals, such long-ranged repulsions between like particles can be attained in experiments by shifting the location of the complementary ``sticky'' DNA sequence from the tail to the middle of the molecule to enhance shell overlap \cite{Mao}. The difference in ranges can be approximately quantified through $\sigma_{\rm AB}/\sigma_{\rm AA}$, which was about 1.5 (square) and 1.2 (BCC) for the more dense crystal structures and about 1.9 for the less dense crystals (Table \ref{tab:converged-variables}). This finding is analogous to previous studies that demonstrated the influence of interaction ranges on DFP self-assembly, for instance, different DNA-linker to particle size ratios producing different structures \cite{PhysRevLett.102.015504, 10.1002/anie.201000633, doi:10.1126/science.1210493}.

We further note that it appeared that some of the interaction parameters would slowly evolve with more iterations of gradient descent, particularly for challenging designs like cubic diamond (Fig.~\ref{figs5}); however, the optimization stopped because the convergence criteria were satisfied. The target $g^{(ij)}$ were well-matched at this point, so this observation supports the notion that multiple acceptable designs may be possible. The design that is found depends on the optimization method, convergence criteria, and initial guess. We will discuss some of these considerations later.

\subsection{Validation of design}
\label{sec:validate}
Inverse design using the relative entropy is sensitive to the simulation protocol because eq.~\eqref{eq:gradS} assumes thermodynamic equilibrium, but equilibrium may not be achieved in a self-assembly simulation for a variety of reasons. For example, a crystal may not nucleate and grow if the simulation time is too short, or there may be crystal defects that cannot anneal. For our simulation protocol, we found that none of the initially crystalline particle configurations were actually able to melt during isochoric temperature cycling despite heating to $T = 3\,\varepsilon/k_{\rm B}$. Lack of melting may indicate thermodynamic stability of the crystal due to a good initial guess of parameters, but it is also possible the crystal is only metastable or the particles are kinetically trapped.

We hence simulated another assembly protocol based on isothermal compression to validate the interactions we designed using isochoric temperature cycling. We placed the same number of particles quasi-randomly in a simulation box whose edge lengths were three times longer than that of the target, then compressed the box isothermally at $T = 1\,\varepsilon/k_{\rm B}$ to its target size by reducing the edge lengths at a constant rate over a period of $5 \times 10^4 \,\tau$. The particles in the expanded box were gas-like, so they were required to assemble in the validation simulations even if the crystal did not melt in the design simulations (Fig.~\ref{figs4}). We then simulated a production period of $10^4\,\tau$, sampling particle configurations every $10^2\,\tau$ for structural analysis. We repeated this procedure 5 times starting from different initial configurations to probe variability.

We are using binary mixtures to assemble the targeted superlattices because the simple interactions we considered are not expected to form them with one particle type; however, from a practical perspective, we may not require perfect compositional order in the final assembled structure. For example, two types of DFPs may have the same nanoparticle core and differ only in their DNA functionalization, making them essentially interchangable in the lattice for properties that depend only on the arrangement of cores. Hence, we employed two different structural metrics that were agnostic to composition to assess if the designed interactions assembled the target structure. The first was based on the radial distribution function $g$ ignoring particle type, which we calculated from the recorded particle configurations up to distance $R = 5\,\sigma$ using bin width $0.05\,\sigma$. We then evaluated the mean squared error (MSE) in $g$ relative to the type-agnostic radial distribution function $g_0$ for the target, 
\begin{equation}
g\:\mathrm{MSE} = \frac{3}{R^3} \int_0^{R} \!\!\! {\rm d}r \, r^2 \left[ g(r) - g_0(r) \right]^2.
\end{equation}
The MSE in $g$ can be nonzero if the particles are correctly arranged in the lattice but have even small differences in their lattice spacing or fluctuations. Accordingly, we chose to use particle-level structural classifiers as a second, complementary metric. We used common neighbor analysis (cutoff radius $1.6\,\sigma$) to classify particles as being in the square lattice \cite{Stukowski_2012, doi:10.1021/acs.langmuir.7b02835}, polyhedral template matching to classify particles as being in the honeycomb---equivalent to graphene---and BCC lattices (root mean square deviation cutoff 0.1) \cite{Larsen_2016}, and an extended common neighbor analysis that identifies second nearest neighbors of a central atom to classify particles as being in the cubic-diamond lattice \cite{MARAS201613}. We then calculated the average fraction of particles that were classified as matching the target structure across all sampled configurations. The analysis was performed using OVITO (version 3.12.0) extended for the square lattice \cite{C8SM00989A}.

\begin{figure}
    \includegraphics{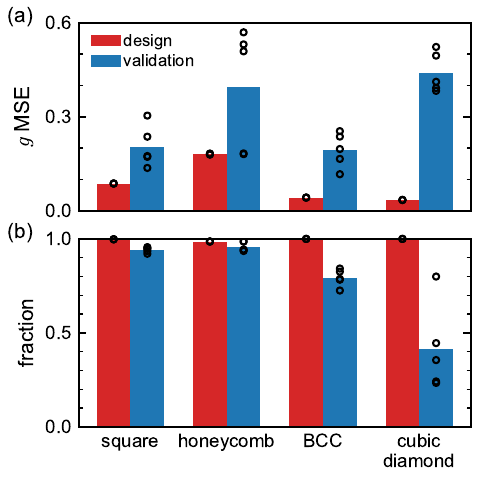}
    \caption{Average (a) mean squared error (MSE) in type-agnostic radial distribution function $g$ and (b) fraction of particles classified as being in target lattice for design (isochoric temperature cycling, red) and validation (isothermal compression, blue) simulation protocols. To probe variability, we performed 3 independent simulations following the design protocol and 5 independent simulations following the validation protocol; the results from each simulation are shown as open circles.}
    \label{fig2} 
\end{figure}

We found that square, honeycomb, and BCC all reliably self-assembled in our validation simulations, with at least 80\% of particles being classified as belonging to the target lattice on average (Fig.~\ref{fig2}). The MSE in $g$ was consistently larger in the validation simulations than in the design simulations. There was also significant variability in $g$ MSE, but not as much in the fraction of particles classified as being in the correct lattice (Fig.~\ref{figs3}). In contrast, there was essentially no variability in either quantity between 3 independent simulations performed using the design protocol, with essentially all particles being classified as the target structure. To better understand the assemblies formed using the validation protocol, we inspected the type-agnostic $g$ for the different crystals (Fig.~\ref{figs2}). We found that although their peaks were typically correctly located, they often had different heights and widths, particularly at longer distances. This behavior is consistent with the presence of defects disrupting long-ranged ordering, which were also visually apparent in particle configurations colored by the structural classifier (Fig.~\ref{figs4}). 

\begin{figure}
    \includegraphics{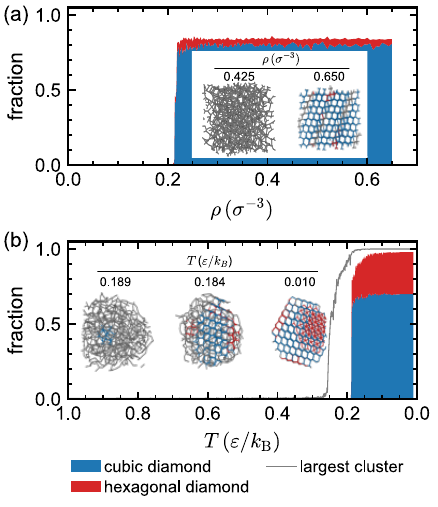}
    \caption{Representative particle snapshots and fractions of particles classified as cubic or hexagonal diamond as a function of (a) number density $\rho$ during bulk isothermal compression and (b) temperature $T$ during isochoric cooling of an initially dilute suspension. In (b), the grey line is the largest cluster size, indicating that essentially all particles have incorporated into the final crystallite. The particle snapshots show the transition from an amorphous cluster to a crystallite with diamond structure. Snapshots here, in Fig.~\ref{fig4}, and in the supplementary information were rendered using OVITO (version 3.12.0) \cite{Stukowski_2012}.}
    \label{fig3} 
\end{figure}

Cubic diamond, on the other hand, had a large increase in $g$ MSE and decrease in fraction of particles classified (only about 27\%) in the validation simulations. There was also significant variability between the 5 independent simulations. The radial distribution functions all showed a lack of long-ranged order, but some also had a few additional small peaks at shorter separation distances that were not present in the target (Fig.~\ref{figs2}d). The structural classification method we used for cubic diamond identified a significant fraction of particles as being hexagonal diamond, a competing polymorph, in some simulations (Fig.~\ref{figs3a}). Despite their substantial structural differences---75\% of rings are boat-like in hexagonal diamond rather than chair-like in cubic diamond---cubic and hexagonal diamond have similar free energies \cite{doi:10.1021/acs.jpcb.0c08723, Romano2012} and subtly different radial distribution functions, making it difficult to select for only one. Overall, our results indicate substantial polymorphism in the diamond structures produced under bulk isothermal compression that was absent under isochoric temperature cycling, likely due to lack of melting. We confirmed that this polymorphism tended to set in early during compression (Fig.~\ref{fig3}a) and persist.

We were curious whether less polymorphism and/or more consistent crystallization might be achievable for cubic diamond by assembling under conditions that more easily allowed for particle rearrangement. In experiments, self-assembly often occurs from a dilute suspension through a gas--solid-like phase transition \cite{doi:10.1073/pnas.2114050118}. Drawing inspiration from this procedure, we initialized simulations in the large simulation box used to start the bulk isothermal compression; however, we now isochorically cooled from $T = 1\,\varepsilon/k_{\rm B}$ to $0.01\,\varepsilon/k_{\rm B}$ at a constant rate over a period of $10^5\,\tau$. This slow cooling should drive the particles to self-assemble finite-size crystallites. Indeed, we found that a single crystallite formed at around $T = 0.18\,\varepsilon/k_{\rm B}$ by a previously identified two-step mechanism \cite{Mao}, in which the particles first aggregated into an amorphous cluster then crystallized (Fig.~\ref{fig3}b). Like the bulk compression simulations, the final crystallite contained both cubic and hexagonal diamond structures; however, unlike the bulk compression simulations (Fig.~\ref{figs3a}), there was significantly less variability in the fraction of each structure between independent simulations (Fig.~\ref{figs9}). The similar free energies of cubic and hexagonal diamond make it challenging to thermodynamically favor one over the other \cite{Mao}. We suspect that our isothermal compression protocol inhibited particles' ability to rearrange after nucleation as they continued to densify, leading to more variability than under dilute assembly conditions.

\subsection{Alternative designs}
To investigate whether acceptable alternatives to the interactions designed in Section \ref{sec:design} may exist, we repeated our inverse-design procedure for the honeycomb lattice with different initial guesses for $\lambda_{\rm AA}$ and $\lambda_{\rm AB}$. We started each of these parameters from either 0 or 1, and we considered all four possible resulting combinations (Fig.~\ref{figs5}). We found that the designed interaction between unlike particles was insensitive to the initial value of either $\lambda_{\rm AA}$ or $\lambda_{\rm AB}$ and was always attractive, but the designed interaction between like particles was dictated by the initial value of $\lambda_{\rm AA}$. Specifically, the designed interaction between like particles was repulsive if $\lambda_{\rm AA} = 0$ initially and attractive if $\lambda_{\rm AA} = 1$ initially. During the design process, $\lambda_{\rm AB}$ stabilized quickly but $\lambda_{\rm AA}$ evolved slowly (Fig.~\ref{figs5}b,d), suggesting the possibility of convergence of $\lambda_{\rm AA}$ across a broad range of values depending on the criteria specified. We will compare the two designs that both initially had $\lambda_{\rm AB} = 1$ and either $\lambda_{\rm AA} = 0$ (repulsive like interactions, designated $-+$) or $\lambda_{\rm AA} = 1$ (attractive like interactions, designated $++$).

\begin{figure}
    \includegraphics{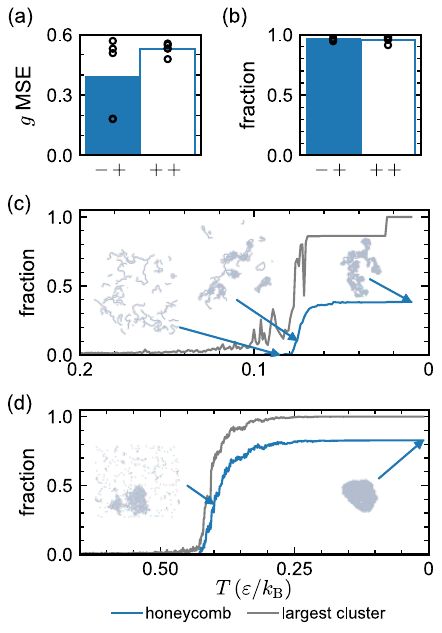}
    \caption{Average (a) mean squared error (MSE) in type-agnostic radial distribution function $g$ and (b) fraction of particles classified as being in target lattice for  5 independent simulations following validation (isothermal compression) protocol using potentials obtained from inverse-design procedure for honeycomb target starting from $(\lambda_{\rm AA}, \lambda_{\rm AB}) = (0, 1)$, denoted $-+$, or $(1, 1)$, denoted $++$. (c, d) Average fraction of particles classified as honeycomb and fraction of particles in largest cluster as a function of temperature $T$ during isochoric cooling of an initially dilute suspension for $-+$ and $++$ designs. Representative snapshots highlighting differences in the assembly process are also shown.}
    \label{fig4} 
\end{figure}

The $-+$ design produced crystals of similar quality as the $++$ design under bulk isothermal compression (Fig.~\ref{fig4}a--b). The $-+$ design had a slightly larger deviation from the target $g$, but there was less variability between independent simulations (Fig.~\ref{fig4}a,  Fig.~\ref{figs6}). Both designs had nearly all particles classified as being in the target structure (Fig.~\ref{fig4}b). However, there were significant differences in assembly from a dilute suspension (Fig.~\ref{figs7}). The $-+$ design required significantly colder temperature to assemble than the $++$ design (Fig.~\ref{fig4}c--d). Assembly occurred in two steps for the $-+$ design, with string-like structures forming before honeycomb, whereas honeycomb directly nucleated for the $++$ design. The assembled structures for the $-+$ design also tended to be less compact than for the $++$ design (Fig.~\ref{figs7}). The structural classifier identified more particles as being honeycomb for the $++$ design than for the $-+$ design; however, this may be partially due to the larger number of interfacial particles in the $-+$ assembly that are more challenging to classify.

The string-like structures for the $-+$ design may be a kinetic effect because the particle dynamics slow considerably as $T$ decreases and assemblies form. To probe this possibility, we ran an additional dilute isochoric simulation where the particles were initialized in the crystallite assembled by the $++$ design, and the temperature was held constant at $T = 0.0755\,\varepsilon/k_{\rm B}$. This temperature is near the nominal melting point for the $-+$ design estimated from a trace of the potential energy during cooling. We ran a long simulation (duration $10^5\,\tau$), but the honeycomb crystallite remained stable and strings did not form. We then ran similar simulations at $0.1 \le k_{\rm B} T / \varepsilon \le 0.5$ at increments of 0.1 (Fig.~\ref{figs11}). We found for the $-+$ potential that strings started to detach from the crystallite surface at $T = 0.1\,\varepsilon/k_{\rm B}$, and the crystal completely melted at higher temperatures. We performed the same test using the $++$ potential and found that it stabilized the crystallite up to at least $T = 0.3\,\varepsilon/k_{\rm B}$. At $T = 0.4\,\varepsilon/k_{\rm B}$, there was a coexistence between the crystal and a dilute phase, before complete melting at $T = 0.5\,\varepsilon/k_{\rm B}$. This behavior is consistent with the crystallization curves observed in cooling simulations (Fig.~\ref{fig4}d).

It has been shown that pair potentials designed for two-dimensional structures can also stabilize their three-dimensional analogs, e.g., a potential designed to self-assemble a square lattice in two dimensions may also self-assemble a simple cubic lattice in three dimensions \cite{Jain2014}. We accordingly tested the ability of both our $-+$ and $++$ designs for honeycomb to form diamond from a dilute suspension. Consistent with our results in two dimensions, we found that the two designs formed distinct structures in three dimensions. In particular, a multilayered graphene-like structure that was not stacked in the manner of graphite was assembled by the $-+$ design (Fig.~\ref{figs8a}), while the $++$ design assembled mostly cubic and hexagonal diamond as well as their closest neighbors (Fig.~\ref{figs8b}). Interestingly for the $-+$ design, the particles first formed strings, as in two dimensions (Fig.~\ref{fig4}c), before the final structure. We suspect that this structure may also be a kinetic effect for similar reasons as in two dimensions. Overall, the difference in transferability to three dimensions demonstrates another practical difference between the two designs, even though both considered were converged and behaved similarly in bulk simulations.

\section{Conclusions}
We have applied relative-entropy minimization to design isotropic pair potentials for binary mixtures of colloidal particles that self-assemble square and honeycomb superlattices in two dimensions as well as BCC and cubic diamond superlattices in three dimensions. By imposing a functional form on the pair potential that is physically motivated by interactions between DFPs, we restrict the design space in a way that forces simpler interactions to be designed and may enhance experimental realizability. Our results show that these simple potentials can still drive assembly of complex lattices, e.g., with low coordination.

A key observation from this work is that multiple viable designs may be identified, depending on the initial parameter guess, but these designs may have rather different behavior under other assembly conditions. To enhance robustness and practical translation, it may then be important to consider the assembly protocol that will be used in experiments when configuring the computational design protocol. For example, dilute simulations of finite-size crystallites might be used for design rather than simulations of bulk crystals. Moreover, while our design strategy assumes equilibrium thermodynamics, designing for nonequilibrium assembly or processing conditions may also be fruitful \cite{D1SM00681A, chatterjeeandjacobs}.

\section*{Author contributions}
BHW and CLP contributed equally to protocol development, data analysis, visualization, and writing the initial draft of the manuscript. QC assisted with data analysis and visualization. BHW conducted simulations. CLP and MPH developed the software used. MPH and JM conceptualized the project, supervised the work, and provided critical feedback throughout. All authors contributed to discussions and editing of the final manuscript.

\section*{Conflicts of interest}
There are no conflicts to declare.

\section*{Data availability}
The software used to perform the inverse design can be found at \\ https://github.com/mphowardlab/relentless. The parameters of the designed potentials are available as supplementary information. The other data that supports the findings of this study are available from the authors upon reasonable request.

\begin{acknowledgments}
Research was sponsored by the Army Research Office and was accomplished under Grant Number W911NF-24-1-0245 (to JM). The views and conclusions contained in this document are those of the authors and should not be interpreted as representing the official policies, either expressed or implied, of the Army Research Office or the U.S. Government. The U.S. Government is authorized to reproduce and distribute reprints for Government purposes notwithstanding any copyright notation herein. CLP was supported by a fellowship from The Molecular Sciences Software Institute under National Science Foundation Award No. CHE-2136142. We gratefully acknowledge the computational resources provided by the Texas A\&M High Performance Research Computing (HPRC) to complete this work.
\end{acknowledgments}

\bibliography{references}

\end{document}

\title{Supplementary information for ``Designing binary mixtures of colloidal particles with simple interactions that assemble complex crystals''
\\ 
}
\author{Beth Hsiao-Yen Wei}
\thanks{These authors contributed equally.}
\affiliation{Artie McFerrin Department of Chemical Engineering, Texas A\&M University, College Station}

\author{C. Levi Petix}
\thanks{These authors contributed equally.}
\affiliation{Department of Chemical Engineering, Auburn University, Auburn, AL 36849, USA}

\author{Qizan Chen}
\affiliation{Artie McFerrin Department of Chemical Engineering, Texas A\&M University, College Station}

\author{Michael P. Howard}
\email{mphoward@auburn.edu}
\affiliation{Department of Chemical Engineering, Auburn University, Auburn, AL 36849, USA}

\author{Jeetain Mittal}
\email{jeetain@tamu.edu}
\affiliation{Artie McFerrin Department of Chemical Engineering, Texas A\&M University, College Station}
\affiliation{Department of Chemistry, Texas A\&M University, College Station}
\affiliation{Interdisciplinary Graduate Program in Genetics and Genomics, Texas A\&M University, College Station}

\maketitle

\begin{table*}[!htbp]
    \caption{\label{tabs1}Lattice vectors (in units of $\sigma$), fractional coordinates and types of particles in unit cell, number of unit cell repeats, and total number of particles for each target crystal structure.}
    \begin{tabular}{l c c c c}
        & square & honeycomb & BCC & cubic diamond \\
        \hline
        lattice vectors &
        $\begin{array}{l}
        \left(2, 0 \right), \\
        \left(0, 2 \right)
        \end{array}$ &
        $\begin{array}{l}
        \left(3, 0 \right), \\
        \left(3/2, 3\sqrt{3}/2 \right)
        \end{array}$ &
        $\begin{array}{l}
        \left(2/\sqrt{3}, 0, 0 \right), \\
        \left(0, 2/\sqrt{3}, 0 \right), \\
        \left(0, 0, 2/\sqrt{3} \right)
        \end{array}$ &
        $\begin{array}{l}
        \left(4/\sqrt{3}, 0, 0 \right), \\
        \left(0, 4/\sqrt{3}, 0 \right), \\
        \left(0, 0, 4/\sqrt{3} \right)
        \end{array}$
        \\ \hline
        unit cell & 
        $\begin{array}{l}
        (0, 0, {\rm A}), \\
        \left(1/2, 0, {\rm B}\right), \\
        \left(0, 1/2, {\rm B}\right), \\
        \left(1/2, 1/2, {\rm A}\right)
        \end{array}$ &
        $\begin{array}{l}
        (0, 0, {\rm A}), \\
        \left(2/3, 0, {\rm B}\right), \\
        \left(0, 1/3, {\rm B}\right), \\
        \left(1/3, 1/3, {\rm A}\right), \\
        \left(1/3, 2/3, {\rm B}\right), \\
        \left(2/3, 2/3, {\rm A}\right)
        \end{array}$ &
        $\begin{array}{l}
        (0, 0, 0, {\rm A}), \\
        \left(1/2, 1/2, 1/2, {\rm B}\right)
        \end{array}$ &
        $\begin{array}{l}
        (0, 0, 0, {\rm A}), \\
        \left(1/2, 1/2, 0, {\rm A}\right), \\
        \left(1/2, 0, 1/2, {\rm A}\right), \\
        \left(0, 1/2, 1/2, {\rm A}\right), \\
        \left(1/4, 1/4, 1/4, {\rm B}\right), \\
        \left(3/4, 1/4, 3/4, {\rm B}\right), \\
        \left(3/4, 3/4, 1/4, {\rm B}\right), \\
        \left(1/4, 3/4, 3/4, {\rm B}\right)
        \end{array}$ \\ \hline
        \# repeats & (16, 16) & (13, 13) & (10, 10, 10) & (5, 5, 5) \\
        total \# particles & 1024 & 1014 & 2000 & 1000
    \end{tabular}
\end{table*}

\clearpage
\begin{figure}[!h]
\includegraphics{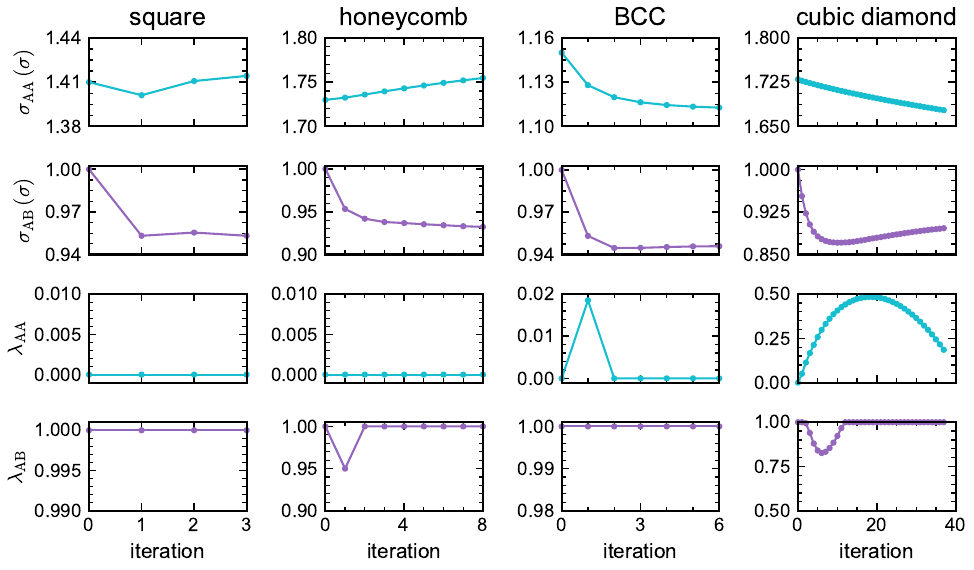}
    \caption{Evolution of variables during optimization for square, honeycomb, BCC and cubic diamond.}
    \label{figs1}
\end{figure}

\begin{table}[!htbp]
\caption{Iteration of optimization at which convergence was achieved and corresponding parameters for each crystal structure.}
\begin{tabular}{lcccc}
 & square & honeycomb & BCC & cubic diamond \\
\hline
iteration & 3 & 8 & 6 & 37 \\
$\sigma_{\mathrm{AA}}$ & 1.414 & 1.755 & 1.113 & 1.677 \\
$\sigma_{\mathrm{AB}}$ & 0.953 & 0.932 & 0.946 & 0.896 \\
$\lambda_{\mathrm{AA}}$ & 0.000 & 0.000 & 0.000 & 0.185 \\
$\lambda_{\mathrm{AB}}$ & 1.000 & 1.000 & 1.000 & 1.000 \\
\end{tabular}
\label{tab:converged-variables}
\end{table}

\begin{figure}[!h]
    \includegraphics{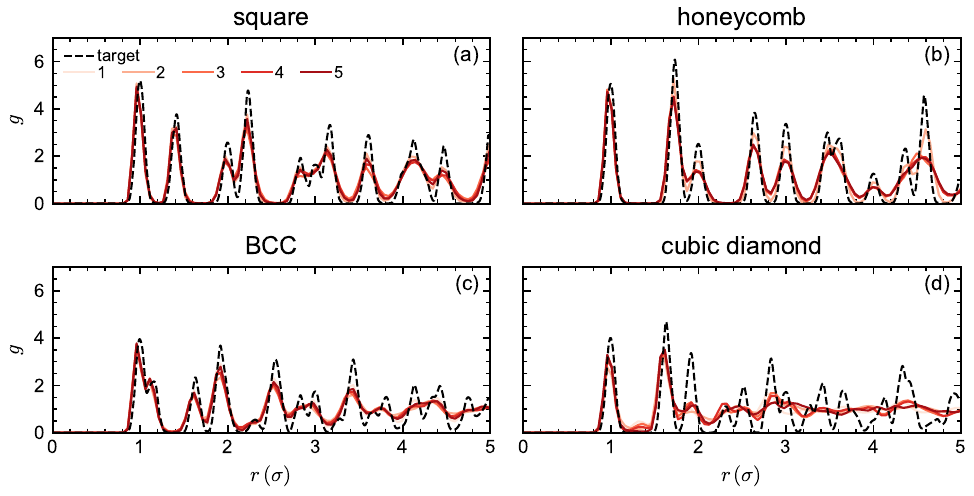}
    \caption{Type-agnostic radial distribution function $g$ sampled from five independent validation (isothermal compression) simulations using the potential designed during isochoric temperature cycling for (a) square (b) honeycomb (c) BCC and (d) cubic diamond.}
    \label{figs2}
\end{figure}

\begin{figure}[!h]
    \includegraphics{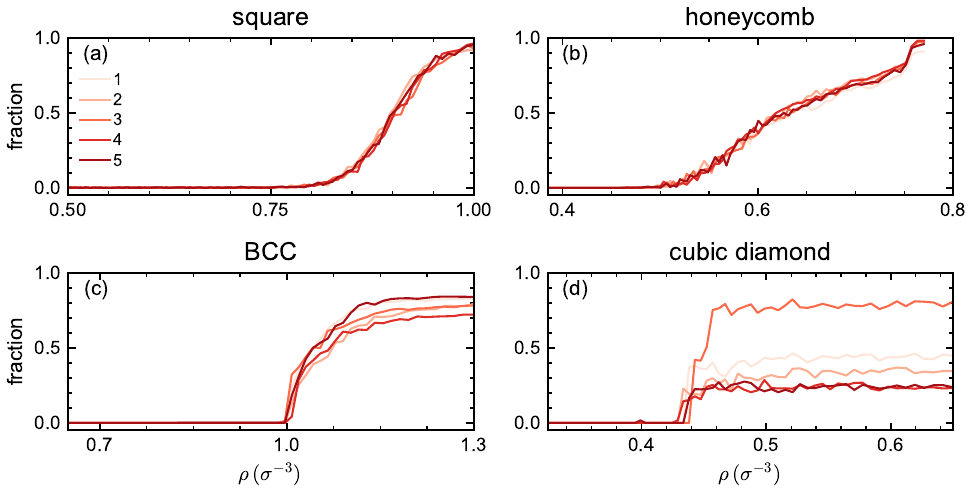}
    \caption{Same as Fig.~\ref{figs2} for the fraction of particles classified as being in target lattice.}
    \label{figs3}
\end{figure}

\begin{figure}[!h]
    \includegraphics{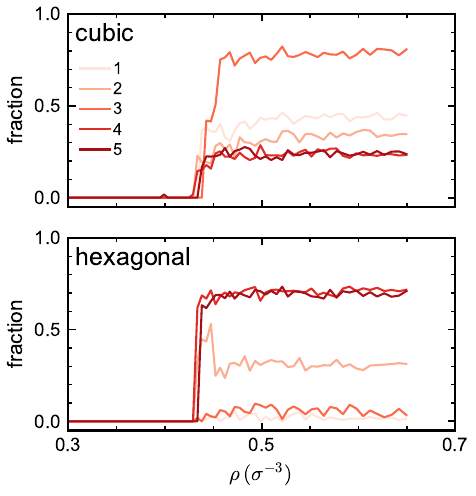}
    \caption{Fraction of particles classified as being cubic or hexagonal diamond for the same simulations as in Figs.~\ref{figs2}d and \ref{figs3}d.}
    \label{figs3a}
\end{figure}

\begin{figure}[!h]
    \includegraphics{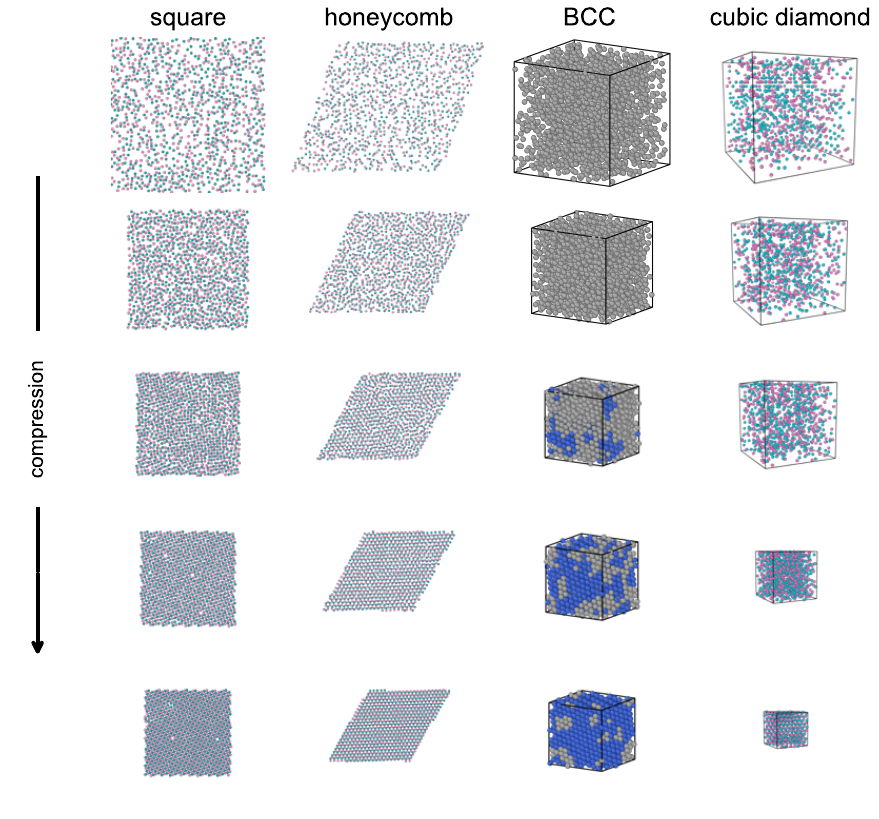}
    \caption{Snapshots from simulation 1 of Fig.~\ref{figs2}.}
    \label{figs4}
\end{figure}

\begin{figure}[!h]
    \includegraphics{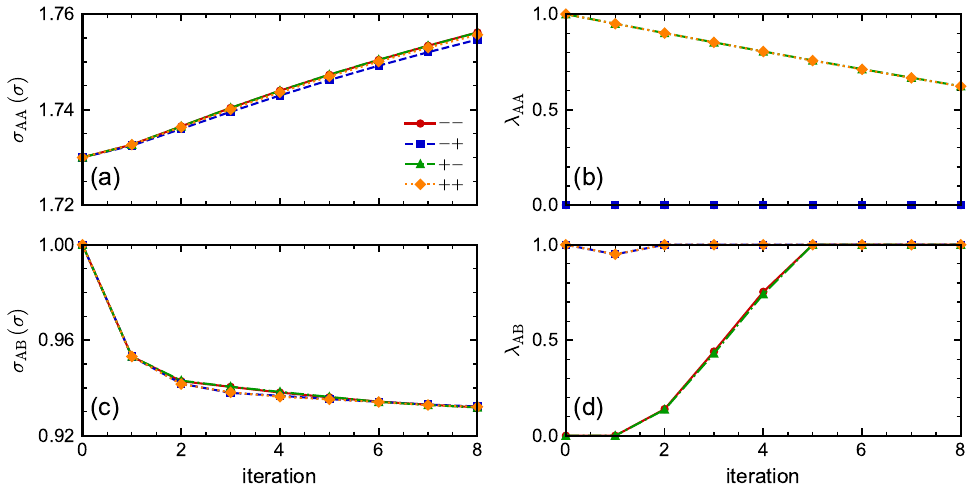}
    \caption{Evolution of variables during optimization for honeycomb initialized from $(\lambda_{\rm AA}, \lambda_{\rm AB}) = (0, 0)$, $(0, 1)$, $(1, 0)$, and $(1, 1)$. The legend uses the notation from the main text, where $-$ is 0 and $+$ is 1. At convergence, the parameters for the $++$ design were $\sigma_{\rm AA} = \,1.756$, $\sigma_{\rm AB} = \,0.932$, $\lambda_{\rm AA} = 0.622$, and $\lambda_{\rm AB} = 1.000$.}
    \label{figs5}
\end{figure}

\begin{figure}[!h]
    \includegraphics{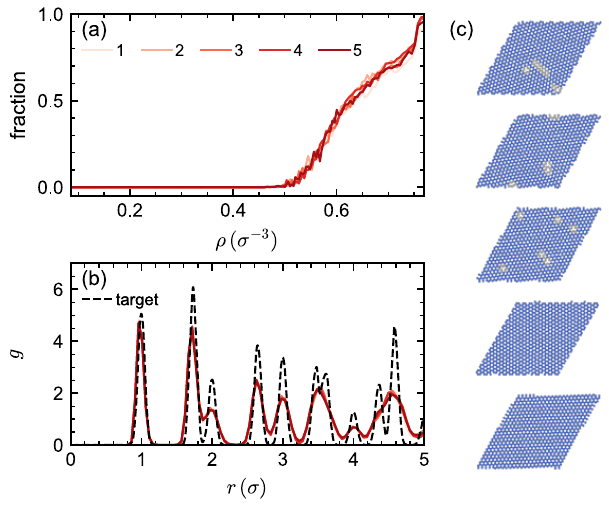}
    \caption{Five independent validation (isothermal compression) simulations for the honeycomb $++$ design. (a) Fraction of particles classified as being honeycomb. (b) Type-agnostic radial distribution function $g$. (c) Final particle configuration. Particles classified as honeycomb are blue, while other particles are gray.}
    \label{figs6}
\end{figure}

\begin{figure}[!h]
    \includegraphics{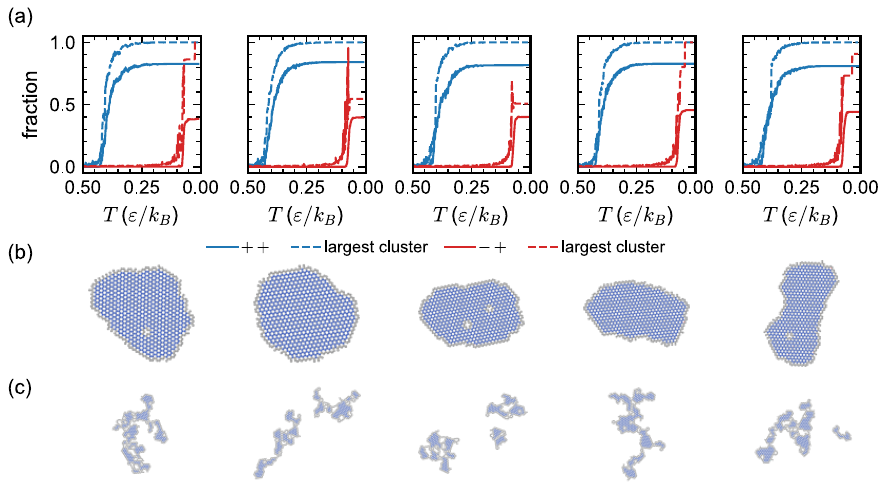}
    \caption{Five independent dilute isochoric cooling simulations in two dimensions for the honeycomb $-+$ and $++$ designs. (a) Fraction of particles classified as being honeycomb and being in the largest cluster. (b) Final particle configuration for the $++$ design. Particles classified as honeycomb are blue, while other particles are gray. (c) Same as (b) for the $-+$ design.}
    \label{figs7}
\end{figure}

\begin{figure}[!h]
    \includegraphics{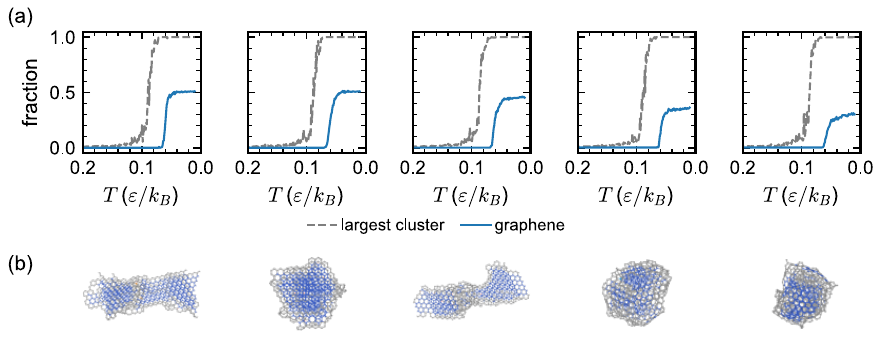}
    \caption{Five independent dilute isochoric cooling simulations in three dimensions for the honeycomb $-+$ design. (a) Fraction of particles classified as being graphene and being in the largest cluster. (b) Final particle configuration. Particles classified as graphene are blue, while other particles are gray.}
    \label{figs8a}
\end{figure}

\begin{figure}[!h]
    \includegraphics{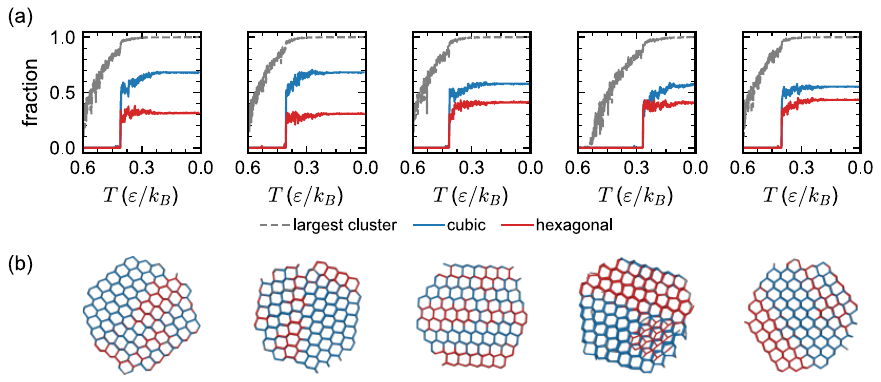}
    \caption{Five independent dilute isochoric cooling simulations in three dimensions for the honeycomb $++$ design. (a) Fraction of particles classified as being cubic or hexagonal diamond and being in the largest cluster. (b) Final particle configuration. Particles classified as cubic diamond are blue or as hexagonal diamond are red, and other particles are gray.}
    \label{figs8b}
\end{figure}

\begin{figure}[!h]
    \includegraphics{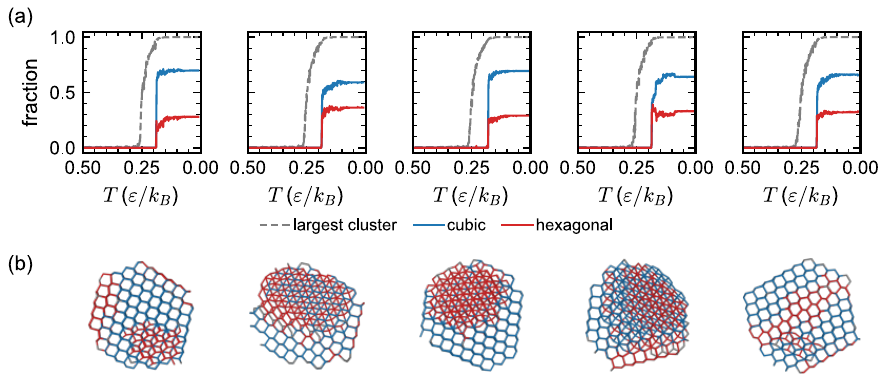}
    \caption{Five independent dilute isochoric cooling simulations for the cubic diamond design. (a) Fraction of particles classified as being cubic and hexagonal diamond and being in the largest cluster. (b) Final particle configuration. Particles classified as cubic diamond are blue or as hexagonal diamond are red, and other particles are gray.}
    \label{figs9}
\end{figure}

\begin{figure}[!h]
    \includegraphics{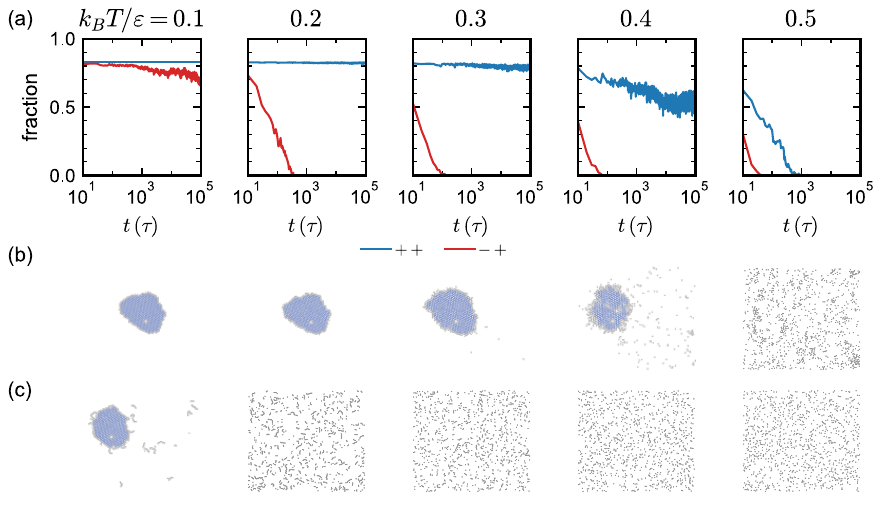}
    \caption{Dilute isothermal--isochoric simulations initialized from the honeycomb crystallite assembled in Fig.~\ref{figs7}b at $0.1 \leq k_BT/\varepsilon \leq 0.5$ in increments of 0.1. (a) Fraction of particles classified as honeycomb for honeycomb $++$ and $-+$ designs. (b) Final particle configuration at each temperature for $++$ design. Particles classified as honeycomb are blue, while other particles are gray. (c) Same as (b) for the $-+$ design.}
    \label{figs11}
\end{figure}